\documentclass{egpubl}
\usepackage{VMV2025}
 
 \WsPaper           %

\usepackage[T1]{fontenc}
\usepackage{dfadobe}  

\usepackage{cite}  %
\BibtexOrBiblatex
\electronicVersion
\PrintedOrElectronic
\ifpdf \usepackage[pdftex]{graphicx} \pdfcompresslevel=9
\else \usepackage[dvips]{graphicx} \fi
\usepackage{amsmath}
\usepackage{amssymb}

\usepackage{egweblnk}

\title[CharGen]{%
  CharGen: Fast and Fluent Portrait Modification
}

\author[J. Dihlmann, A. Killguss, H. Lensch]{%
  Jan-Niklas Dihlmann$^{1,2}$\orcid{0009-0007-8279-229X} 
  Arnela Killguss$^{1,2}$\orcid{0009-0008-3340-1798}
  Hendrik Lensch$^{1}$\orcid{0000-0003-3616-8668} \\
  $^1$University of Tübingen, Germany \\
  $^2$Equal Contribution
}

\usepackage{booktabs,siunitx,xcolor,graphicx,colortbl}
\usepackage{tabularx}

\definecolor{bestcol}{RGB}{254,196,79}      %
\definecolor{secondcol}{RGB}{255,247,188}   %

\newcommand{\best}[1]{\cellcolor{bestcol}\bfseries #1}
\newcommand{\secondbest}[1]{\cellcolor{secondcol} #1}

\begin{document}

\teaser{
    \includegraphics[width=0.9\linewidth]{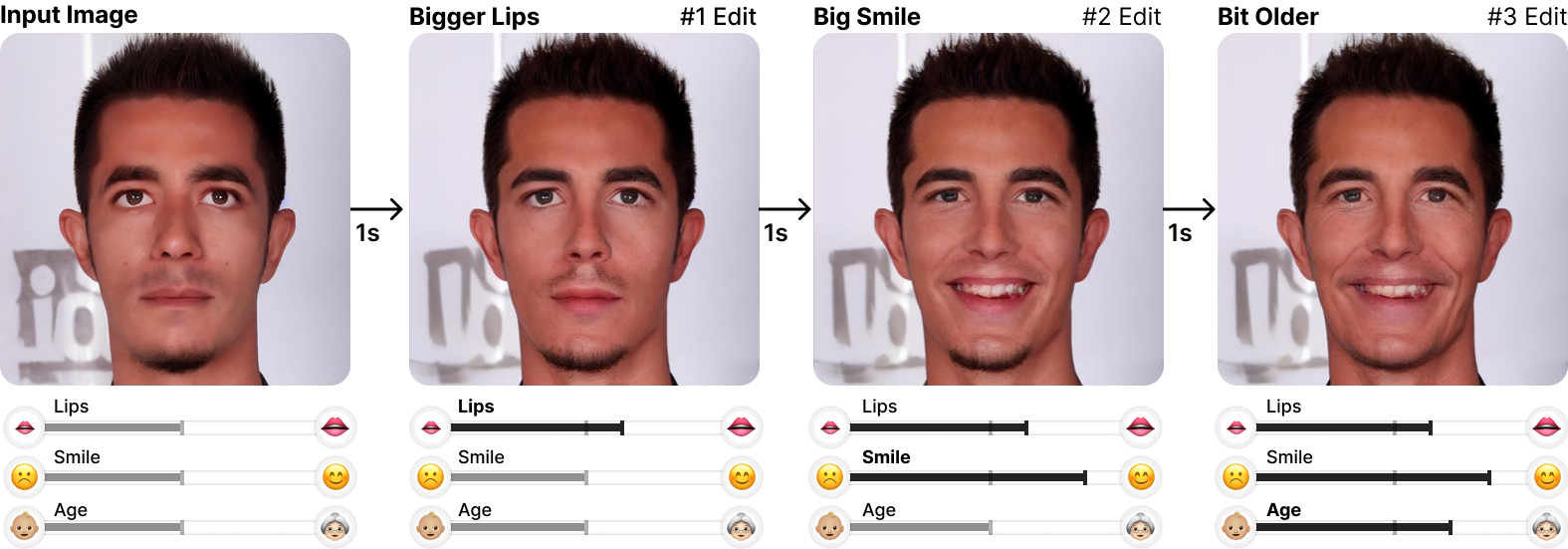}
    \centering
    \caption{CharGen enables fast, interactive editing of character images with continuous sliders for multiple facial attributes.}
    \label{fig:teaser}
}

\maketitle

\begin{abstract}
    Interactive editing of character images with diffusion models remains challenging due to the inherent trade-off between fine-grained control, generation speed, and visual fidelity.
    We introduce CharGen, a character-focused editor that combines attribute-specific Concept Sliders, trained to isolate and manipulate attributes such as facial feature size, expression, and decoration with the StreamDiffusion sampling pipeline for more interactive performance.
    To counteract the loss of detail that often accompanies accelerated sampling, we propose a lightweight Repair Step that reinstates fine textures without compromising structural consistency.
    Throughout extensive ablation studies and in comparison to open-source InstructPix2Pix and closed-source Google Gemini, and a comprehensive user study, CharGen achieves two-to-four-fold faster edit turnaround with precise editing control and identity-consistent results. 
    Project page: \url{https://chargen.jdihlmann.com/}

\end{abstract}

\begin{section}{Introduction}
\label{sec:introduction}

Character-focused 3D editors like The Sims 4~\cite{thesims4} or Unreal MetaHumans~\cite{metahumans} and 2D editors like Character Creator~\cite{charactercreator} and CC2D~\cite{cc2d} offer intuitive sliders and drag-and-drop controls for predefined attributes. While these systems provide excellent control over character creation and editing, they require handcrafted attribute definitions and engine-specific rigging, which can limit their flexibility for rapid prototyping or editing arbitrary portraits without some setup time.
Designing such systems demands expertise to predefine editable attributes, though modern 3D morphable models offer good generalization to new identities through their statistical foundations.

Recent diffusion-based models (e.g., Stable Diffusion~\cite{rombach2022highresolutionimagesynthesislatent}), deliver high-fidelity synthesis from text or image prompts, yet their base architectures lack mechanisms for precise, continuous attribute editing.
Instruction-driven editing methods like open-source InstructPix2Pix~\cite{brooks2023instructpix2pixlearningfollowimage} or Flux Kontext~\cite{labs2025flux1kontextflowmatching}.
Closed-source systems like Google Gemini~\cite{geminiteam2024geminifamilyhighlycapable} allow semantic modifications but suffer from discrete instructions, long single-edit generation times, and user-unfriendly pipelines for sequential edits.
Hybrid tools (e.g., Photoshop Firefly~\cite{adobe_photoshop, adobe_firefly}) integrate masking with generative fills, but masking remains laborious and imprecise, and editing speed lags in iterative workflows.
Early work on Concept Sliders~\cite{gandikota2023conceptslidersloraadaptors} demonstrated continuous latent-space control of attributes such as facial structure, expression, and style via LoRA adapters, but is limited by standard sampling speeds.

In this work, we present CharGen, a continuous generative character editing pipeline that (1) harnesses attribute-specific Concept Sliders, trained via our reproducible recipe, (2) integrates the StreamDiffusion sampler for two to fourfold faster edits~\cite{kodaira2023streamdiffusionpipelinelevelsolutionrealtime}, and (3) introduces a lightweight, identity-preserving Repair Step to restore high-frequency detail lost during accelerated sampling.

CharGen empowers artists and developers with an out-of-the-box tool for rapid, interactive portrait editing, reducing manual masking and iteration time while preserving creative control and visual fidelity.
Developers can also adapt our pipeline for different continuous attribute editing tasks in various domains such as item variation and scene composition. 
\end{section}

\begin{section}{Related Work}
\label{sec:related_work}

Traditional 2D editing tools like Adobe Photoshop~\cite{adobe_photoshop} and Canva~\cite{canva} provide pixel-level control through masking, warping, and color grading.
These deliver precise edits but require significant skill and time for sequential modifications.
Early generative models like PixelCNN and PixelRNN~\cite{oord2016pixelrecurrentneuralnetworks} modeled images pixel-by-pixel for coherent outputs.
Generative Adversial Networks (GANs)~\cite{goodfellow2014generativeadversarialnetworks} improved realism through adversarial training between generator and discriminator.
SliderGAN~\cite{ververas2019slidergan} demonstrated the potential of combining 3D morphable models with GANs for continuous parameter-based face editing, showing how blendshape parameters could control facial expressions in generated images.
Recent hybrid Autoregressive-Diffusion frameworks~\cite{tang2024hartefficientvisualgeneration} combine sequential prediction with diffusion sampling, while Diffusion Vision Transformers~\cite{hatamizadeh2024diffitdiffusionvisiontransformers} integrate transformer backbones directly into diffusion processes.

Generative editing methods can be categorized by their input modalities.
In-painting and sketch-guided approaches let users define regions for model refinement.
SDEdit~\cite{meng2022sdeditguidedimagesynthesis} uses rough sketches to guide diffusion sampling, and SOEDiff~\cite{wu2024soediffefficientdistillationsmall} distills diffusion priors for efficient text-prompted in-painting.
These reduce manual masking but lack continuous control over specific attributes and often need multiple refinement steps.
Point-dragging interfaces translate gestures into geometric transformations.
DragGAN~\cite{Pan_2023} enables interactive deformation in GAN latent spaces, and DragDiffusion~\cite{shi2024dragdiffusionharnessingdiffusionmodels} adapts this to diffusion models for semantically consistent operations.
These excel at shape manipulation but don't extend to attribute modulation like expression or style.
Text-driven editing interprets natural language prompts to guide modifications.
DiffusionCLIP~\cite{kim2022diffusioncliptextguideddiffusionmodels} fuses CLIP guidance with diffusion sampling for style transfer and object alteration.
InstructPix2Pix~\cite{brooks2023instructpix2pixlearningfollowimage} trains diffusion models to follow free-form editing commands, and commercial systems like Google Gemini~\cite{geminiteam2024geminifamilyhighlycapable} offer end-to-end text-based pipelines.
These are accessible but operate in discrete steps with noticeable latency.

To achieve continuous semantic editing, fine-grained control, slider-based interfaces have emerged.
Prompt Sliders~\cite{sridhar2024promptslidersfinegrainedcontrol} learn embedding directions via Textual Inversion for composable attribute adjustments with minimal overhead.
Concept Sliders~\cite{gandikota2023conceptslidersloraadaptors} train LoRA adapters to isolate latent directions for specific attributes.
Prompt Sliders minimize inference overhead but suffer from attribute entanglement; Concept Sliders deliver precise control but inherit slower edit turnaround.

Real-time generative editing research targets sub-second feedback for interactive workflows.
Imagic~\cite{kawar2023imagictextbasedrealimage} distills inversion techniques for near-instant text-based modifications, and StyleMapGAN~\cite{kim2021exploitingspatialdimensionslatent} exploits spatial style modulation within GANs for on-the-fly local edits.
These achieve impressive speeds but remain tied to GAN domains or support limited attribute ranges.

In summary, existing systems trade off between attribute granularity, latency, and visual fidelity when editing arbitrary character images.
CharGen bridges these gaps by combining attribute-specific Concept Sliders with StreamDiffusion sampling~\cite{kodaira2023streamdiffusionpipelinelevelsolutionrealtime} for two-to-fourfold faster edits, plus a lightweight Repair Step that restores high-frequency details without compromising structural consistency.
\end{section}

\begin{section}{Background}
    \label{sec:background}

    \begin{paragraph}*{Diffusion Models}
    \cite{sohldickstein2015deepunsupervisedlearningusing} learn to gradually denoise images through forward and backward diffusion processes.
    Latent Diffusion Models (LDMs)~\cite{rombach2022highresolutionimagesynthesislatent} operate in compressed latent space rather than pixel space, significantly reducing computational complexity.
    Stable Diffusion (SD)~\cite{rombach2022highresolutionimagesynthesislatent} uses a Variational Autoencoder (VAE) to encode images $x$ into latent representations $z = \mathcal{E}(x)$ and decode them back $x' = \mathcal{D}(z)$.
    The UNet architecture with skip connections handles the denoising process, predicting noise $\epsilon_{\theta}(z_t, t)$ at each timestep $t$.
    Classifier-Free Guidance (CFG)~\cite{ho2022classifierfreediffusionguidance} ensures text prompt adherence by computing positive and negative conditioning terms.
    \end{paragraph}

    \begin{paragraph}*{LoRA Fine-tuning}
    Full fine-tuning of LDMs requires updating all model weights, which is computationally expensive.
    Low-Rank Adapters (LoRA)~\cite{hu2021loralowrankadaptationlarge} freeze the original model and inject trainable rank decomposition matrices.
    Instead of modifying the weight matrix $W_0 \in \mathbb{R}^{d \times k}$, LoRA introduces low-rank decomposition:
\begin{equation}
W' = W_0 + \Delta W = W_0 + BA
\label{eq:loraweight}
\end{equation}
    where $A \in \mathbb{R}^{r \times k}$ and $B \in \mathbb{R}^{d \times r}$ are trainable adapters.
    Only parameters $A$ and $B$ need updating, making task switching efficient by recovering $W_0$ and using different $A'$ and $B'$.
    \end{paragraph}

    \begin{paragraph}*{Concept Sliders}
    Gandikota et al.\ \cite{gandikota2023conceptslidersloraadaptors} enable precise attribute control by fine-tuning LoRAs for specific concepts.
    The method learns parameter directions that increase target attribute $c_+$ while decreasing negative attribute $c_-$:
\begin{equation}
\epsilon_{\theta^*}(X, c_t, t) \leftarrow \epsilon_{\theta}(X, c_t, t) + 
\eta (\epsilon_{\theta}(X, c_+, t) - \epsilon_{\theta}(X, c_-, t))
\label{eq:diffusion}
\end{equation}
Disentanglement ensures modifications are consistent across preservation attributes $p$:
\begin{equation}
\epsilon_{\theta^*}(X, c_t, t) \leftarrow \epsilon_{\theta}(X, c_t, t) + \eta \sum_{p \in P} \Delta\epsilon_p
\label{eq:disentanglement}
\end{equation}
where $\Delta\epsilon_p = \epsilon_{\theta}(X, (c_+, p), t) - \epsilon_{\theta}(X, (c_-, p), t)$.
Textual Concept Sliders use text prompt pairs, while Visual Concept Sliders require image datasets with positive/negative examples.
The scaling factor $\alpha$ controls attribute strength without retraining.
    \end{paragraph}

    \begin{paragraph}*{StreamDiffusion}
    \cite{kodaira2023streamdiffusionpipelinelevelsolutionrealtime} achieves real-time generation through several optimizations.
    Stream Batch processes multiple denoising steps in a single UNet pass, reducing the total inference count.
    Residual Classifier-Free Guidance (RCFG) approximates negative conditions with virtual residual noise, avoiding additional UNet passes.
    Input-Output Queues handle tensor conversion outside the neural pipeline for parallel processing.
    The Stochastic Similarity Filter skips denoising for similar input images using cosine similarity thresholds.
    Pre-computation caches prompt embeddings, Key-Value pairs, Gaussian noise, and noise strength coefficients.
    Model acceleration uses TensorRT and TinyAutoEncoder for faster latent space transformation.
    These optimizations enable real-time generation but introduce quality trade-offs that CharGen addresses through its Repair Step.
    \end{paragraph}
\end{section}

\begin{section}{Method}
\label{sec:method}

CharGen combines attribute-specific Concept Sliders with StreamDiffusion's real-time pipeline to enable interactive character editing.
The system addresses three key challenges: (1) fine-grained attribute control through pretrained Concept Sliders, (2) interactive inference speed via StreamDiffusion integration, and (3) detail restoration through a lightweight Repair Step.

\subsection{Attribute-Specific Concept Sliders}
We pretrain Concept Sliders for four critical character attributes: facial expressions (smile, surprise, anger), facial structure (eye size, lip size, weight), aging and stylization, and hair adjustments.
Each slider learns directional shifts in latent space using the optimization described in Section \ref{sec:background}, enabling precise attribute modification while preserving character identity.
Training employs both textual prompts and paired image datasets we use Flickr-Faces-HQ (FFHQ)~\cite{karras2019stylebasedgeneratorarchitecturegenerative}, where each pair represents controlled single-attribute modifications.
The resulting sliders provide continuous control over specific attributes without affecting others, crucial for maintaining character consistency during interactive editing. 
Due to their independence multiple concept sliders can easily be combined and applied at the same time by simply averaging the corresponding LoRA weight matrices.

\subsection{StreamDiffusion Integration}
We integrate pretrained Concept Sliders into StreamDiffusion's pipeline through LoRA merging, avoiding additional fine-tuning while maintaining fast performance.
We explored two approaches for combining Concept Sliders with StreamDiffusion: LoRA stacking and LoRA merging.
LoRA stacking uses the existing $load\_lora$ and $fuse\_lora$ functions to progressively apply multiple LoRAs one after another during the generation process, where each slider is loaded and fused sequentially at runtime.
In contrast, LoRA merging uses the $set\_adapters$ function to pre-combine multiple Concept Sliders by summing their weight matrices before injection into the model.
While stacking can offer faster loading times in some implementations, it introduces limitations in the StreamDiffusion pipeline that we analyze in Section \ref{sec:experiments}.
We therefore adopt LoRA merging as our primary method, which provides more stable and efficient multi-attribute control.

Given a set of Concept Sliders $(W_1, W_2, ..., W_k)$ with scaling factors $\alpha_i$, the combined LoRA update is computed as:
\begin{equation}
W_{\text{merged}} = W_0 + \sum_{i=1}^{k} \alpha_i W_i
\label{eq:merged}
\end{equation}
The merged weights are injected into StreamDiffusion's UNet using the $set\_adapters$ function, enabling dynamic attribute adjustment through slider coefficients.
The pipeline processes input images through four batched denoising steps, encoding via TinyAutoEncoder, applying Concept Slider transformations, and decoding back to pixel space.
This approach enables simultaneous multi-attribute modification while maintaining StreamDiffusion's efficiency.
Figure \ref{fig:pipeline} illustrates the complete pipeline, showing how individual LoRA matrices are merged and injected into the generation process.

\begin{figure*}[htb]
    \centering
    \includegraphics[width=\linewidth]{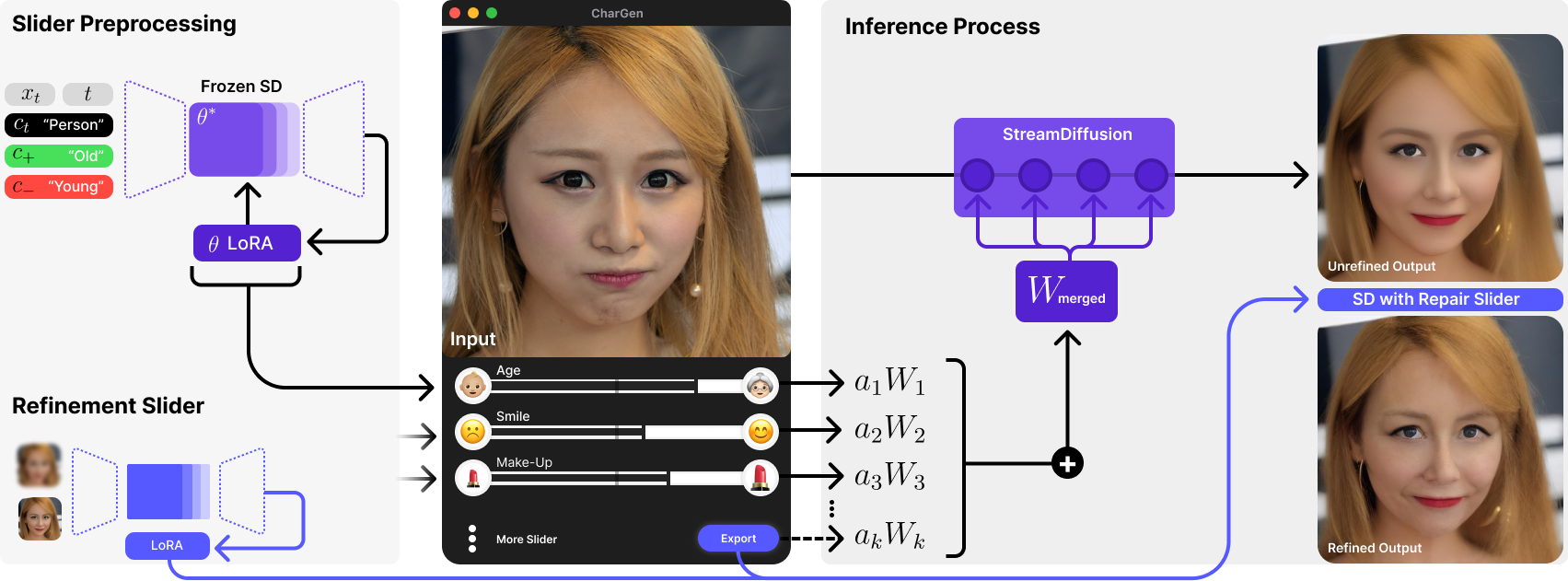}
    \caption{Overview of the CharGen pipeline. The pipeline is based on Concept Sliders~\cite{gandikota2023conceptslidersloraadaptors} and StreamDiffusion~\cite{kodaira2023streamdiffusionpipelinelevelsolutionrealtime} and uses LoRA merging to combine multiple Concept Sliders. (Left) Pretraining of Concpet Sliders is achieved via Text or Image to Image pairs. (Center) The user can adjust the sliders to control the attributes of the generated image. (Right) During inference, the Concept Sliders are merged with the base model given the selected slider strength and the image is generated. The user can make multiple edits by adjusting the sliders. (Bottom) The user can export the image and the image is then refined using the Repair Slider approach. }
    \label{fig:pipeline}
\end{figure*}

\subsection{Repair Step}
StreamDiffusion's acceleration introduces detail loss that we address through a lightweight Repair Step.
We explore three approaches: (A) standard Stable Diffusion, (B) Repair Slider integration, and (C) ControlNet-based repair.
We focus on methods B and C, deferring comparative analysis to Section \ref{sec:experiments}.

\begin{paragraph}*{Repair Slider Training}
We train a specialized Repair Slider to restore details lost during StreamDiffusion processing.
The training dataset pairs high-resolution ground truth images with their StreamDiffusion-processed counterparts, teaching the slider to move from low-detail outputs toward realistic, detailed versions while preserving structural consistency.
\end{paragraph}

\begin{paragraph}*{ControlNet Repair}
Second, we adapt ControlNet LoRA~\cite{wu2024controllorav3} for the repair task, leveraging its structural guidance capabilities.
Training uses the same high-resolution dataset but introduces identity-preserving variations through InstantID~\cite{wang2024instantid} perspective shifts and subtle makeup modifications.
\end{paragraph}

The complete CharGen pipeline enables interactive character editing with precise attribute control and detail preservation, addressing the fundamental trade-offs between speed, control, and visual fidelity in interactive generative editing.
\end{section}

\begin{section}{Experiments}
    \label{sec:experiments}

    We evaluate CharGen's effectiveness through qualitative and quantitative comparisons against existing editing methods, including InstructPix2Pix~\cite{brooks2023instructpix2pixlearningfollowimage} and Google Gemini~\cite{geminiteam2024geminifamilyhighlycapable}.
    Our evaluation focuses on three key aspects: method selection and implementation details,  qualitative comparison of editing capabilities, and quantitative assessment of performance and fidelity. 
    We also conduct a user study to validate perceptual quality across different editing scenarios.

    \subsection{Implementation and Method Selection} 
    \label{sec:implementation}
    We integrate Concept Sliders~\cite{gandikota2023conceptslidersloraadaptors} within the StreamDiffusion~\cite{kodaira2023streamdiffusionpipelinelevelsolutionrealtime} framework using their real-time image-to-image demo.
    Our Concept Sliders are trained on Stable Diffusion 1.4~\cite{rombach2022highresolutionimagesynthesislatent} with scaling factors from -10 to 10 for interactive control.
    Training utilizes the Flickr-Faces-HQ (FFHQ)~\cite{karras2019stylebasedgeneratorarchitecturegenerative} dataset for facial attribute modifications.
    All experiments are conducted on an NVIDIA RTX 3090 GPU with 24GB VRAM.

    \begin{figure}
\centering
\includegraphics[width=\columnwidth]{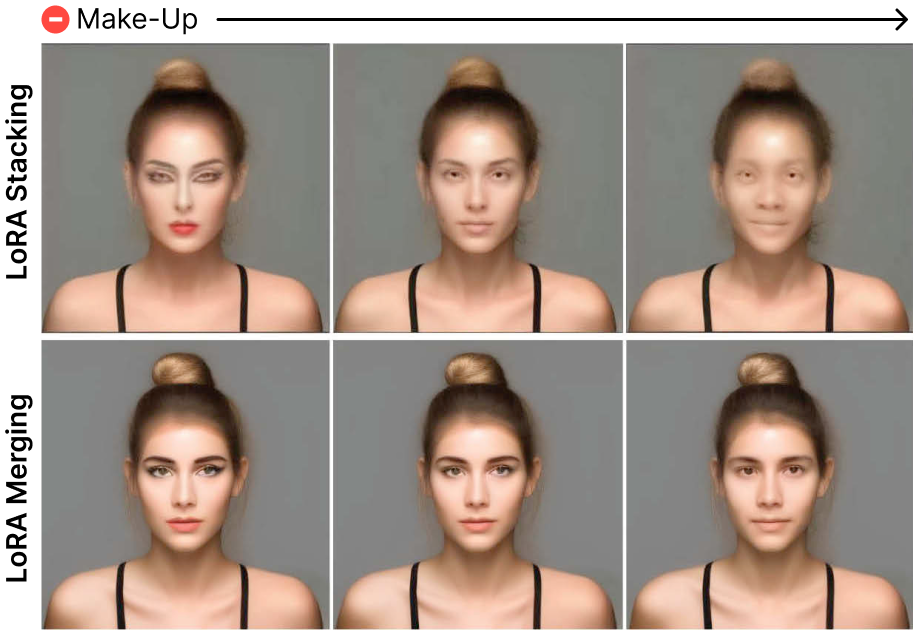}
\caption{
Visual comparison of LoRA stacking and LoRA merging for progressive make-up removal w/o repair. 
(Top) LoRA stacking applies each Concept Slider sequentially at runtime, which leads to cumulative distortions and loss of facial detail as more multiple changes or sliders are applied. 
(Bottom) LoRA merging precombines all Concept Sliders by summing their weight matrices, resulting in stable and consistent attribute changes without progressive degradation. 
}
\label{fig:lora_methods}
\end{figure}

    \begin{paragraph}*{LoRA Integration Strategy}
    Initial experiments revealed that dynamic LoRA stacking during inference caused progressive distortions due to weight accumulation, as illustrated in Figure~\ref{fig:lora_methods}. 
    Stacking LoRAs sequentially led to a visible loss of facial detail and unnatural artifacts when multiple sliders were applied in succession. 
    To address this, we adopt LoRA merging, which combines multiple Concept Sliders while maintaining independent scaling factors. 
    This approach ensures stability and consistency across sequential modifications, enabling continuous multi-attribute editing without quality degradation.
    \end{paragraph}

    \begin{paragraph}*{Refinement Method Selection}
    Comparative analysis of our three refinement approaches shows that ControlNet~\cite{zhang2023addingconditionalcontroltexttoimage} produces excessive structural deviations from ground truth.
    Standard Stable Diffusion provides moderate detail but lacks sharpness in fine features.
    Our Repair Slider approach offers the best balance of detail enhancement and structural preservation, making it the preferred choice for subsequent evaluations (Figure \ref{fig:refinement_comparison}).
    \end{paragraph}

    \begin{figure*}
\centering
\includegraphics[width=\textwidth]{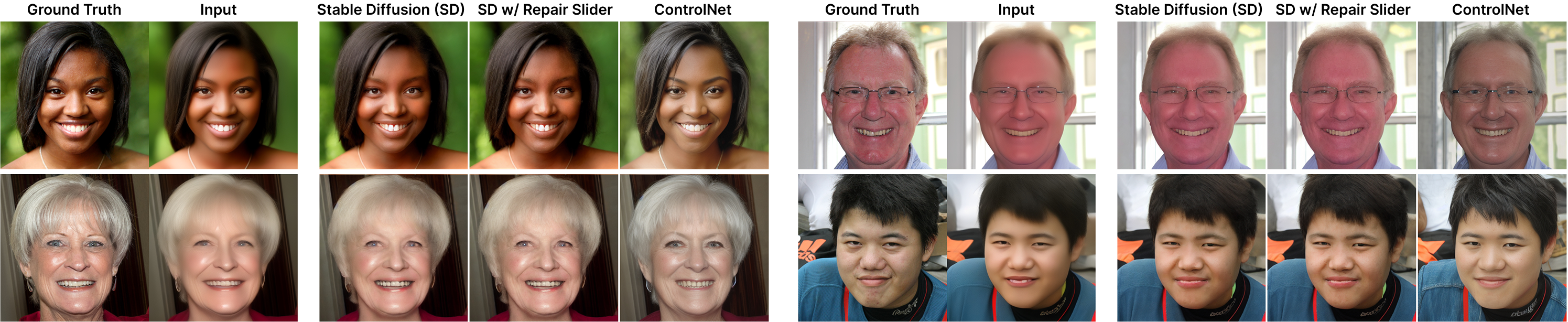}
\caption{Comparison of our three Refinement Steps Standard Stable Diffusion, Repair Slider and ControlNet. The Input image is the Ground Truth image after passing through StreamDiffusion once. Showcasing that the Repair Slider is the best balance of detail enhancement and structural preservation.}
\label{fig:refinement_comparison}
\end{figure*}

    \subsection{Qualitative Evaluation}
    We compare CharGen against InstructPix2Pix and Gemini across four attribute categories: facial expressions, structural modifications, aging effects, and hair adjustments.
    Evaluation criteria include visual accuracy, identity preservation, and editing consistency.

    \begin{figure*}
\centering
\includegraphics[width=\linewidth]{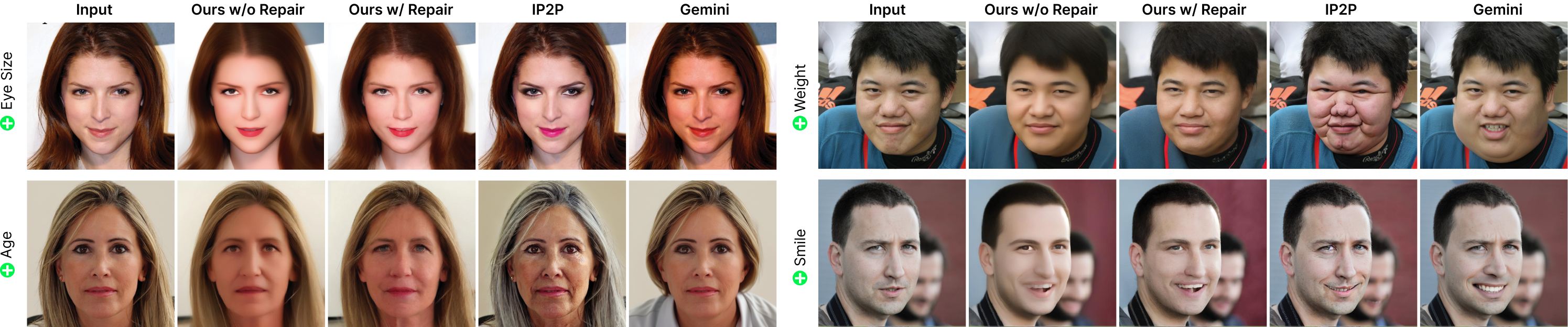}
\caption{Single attribute modifications. Comparison of CharGen against InstructPix2Pix and Gemini for different facial attribute modifications. Plus/minus symbols indicate modification direction (e.g., "Age +" indicates aging effect). Find prompts and scales in the supplementary.}
\label{fig:single_concepts}
\end{figure*}

    \begin{paragraph}*{Single Attribute Modifications}
    InstructPix2Pix excels at major structural transformations but struggles with fine-grained control, requiring prompt engineering for subtle modifications (Figure \ref{fig:single_concepts}).
    Gemini demonstrates consistent performance across attribute types but occasionally fails due to content moderation constraints.
    CharGen provides precise control for localized adjustments like makeup and facial features, though it shows limitations with strong transformations such as aging effects.
    Unlike existing methods that provide discrete editing steps, CharGen is the only approach enabling truly continuous control through progressive slider adjustments. Figure~\ref{fig:lora_methods} demonstrates how makeup removal can be smoothly transitioned across intermediate values, providing fine-grained control that is essential for iterative editing workflows. A supplementary video further illustrates this continuous control capability across various attributes.
    \end{paragraph}

    \begin{paragraph}*{Multi-Attribute Editing}
    CharGen's LoRA merging approach enables simultaneous modification of multiple attributes while maintaining consistency (Figure \ref{fig:multiple_concepts}).
    InstructPix2Pix and Gemini struggle to incorporate all requested changes simultaneously, often prioritizing dominant attributes.
    Our method's interactive nature allows interactive adjustment of multiple sliders, providing immediate visual feedback for complex editing scenarios.
    \end{paragraph}

    \begin{figure*}
\centering
\includegraphics[width=\linewidth]{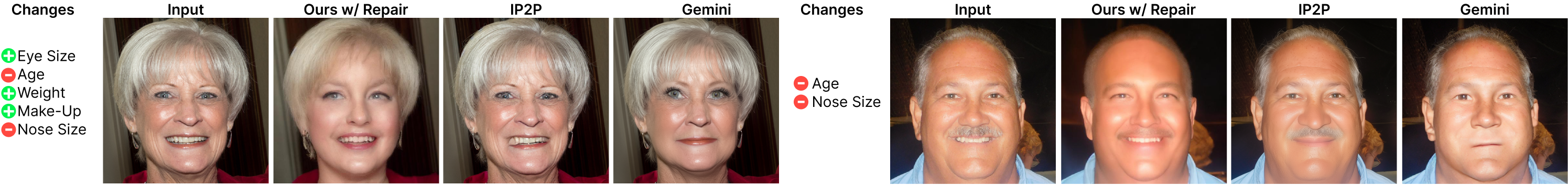}
\caption{Examples for modifying multiple facial attributes at once. The attributes are listed with the direction of change. Our edit is visually closer to the desired changes while the other two methods don't manage to incorporate all desired changes or modify other attributes.}
\label{fig:multiple_concepts}
\end{figure*}

    \begin{paragraph}*{Progressive Editing}
    Sequential editing with InstructPix2Pix and Gemini accumulates artifacts due to successive image processing (Figure \ref{fig:progressive_concepts}).
    CharGen maintains the original input throughout the editing process by only adjusting slider parameters for progressive modifications.
    This approach eliminates quality degradation while enabling fluent attribute transitions.
    \end{paragraph}

    \begin{figure*}
\centering
\includegraphics[width=\linewidth]{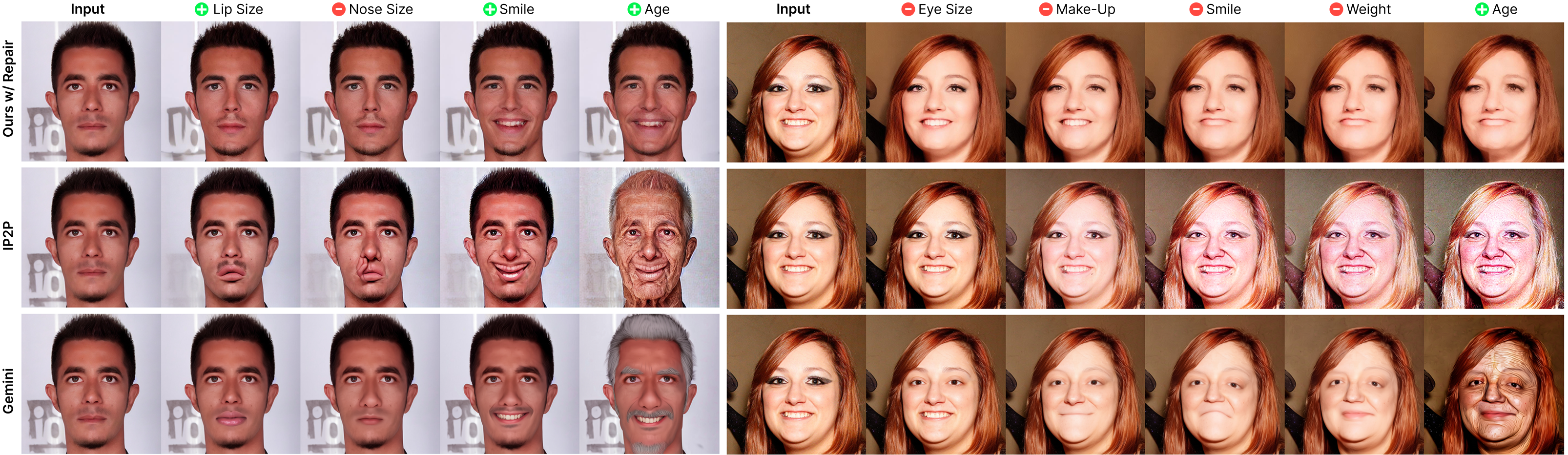}
\caption{Comparison of Gemini, IP2P and our method for progressive edits (left to right). Our method succeeds in fluent modifications while maintaining previous edits. The Input stays the same for our method, while IP2P and Gemini use the Input initially and then the previous edit.}
\label{fig:progressive_concepts}
\end{figure*}

    \begin{paragraph}*{Concept Slider Interactions}
    While LoRA merging enables multi-slider usage, some combinations exhibit interference effects.
    For example, age modifications affect lip size due to correlated anatomical changes, and makeup-age combinations can reduce image sharpness.
    These effects arise from independent slider training without cross-attribute awareness, highlighting opportunities for future joint optimization approaches. 
    \end{paragraph}

    \subsection{Quantitative Analysis}
    We evaluate performance across three metrics: generation speed, modification strength, and image fidelity.

    \begin{paragraph}*{Generation Speed}
    Our system achieves $2.55$ seconds with $14$ active LoRAs, compared to $6.62$ seconds for Gemini and $33.00$ seconds for InstructPix2Pix (Table \ref{tab:time_comparison}).
    While not strictly real-time, our system provides interactive response times suitable for iterative editing workflows, which depends on the amount of LoRAs.
    \end{paragraph}

    \begin{table}[!h]
    \centering
    \scriptsize
    \setlength{\tabcolsep}{4pt} %

    \begin{tabularx}{\linewidth}{l *{4}{>{\centering\arraybackslash}X}}
      \toprule
      \textbf{Method Comparison} & & & & \\[-1.1em]
      & \textbf{Ours} & \textbf{Gemini$^*$} & \textbf{IP2P} & \\
      \midrule
      \textit{Average time (seconds)} & \best{2.55} & \secondbest{6.62} & 33.00 & \\
      \midrule
      \midrule
      \textbf{Amount of LoRAs / Attributes} & & & & \\[-1.1em]
      & \textbf{1} & \textbf{5} & \textbf{8} & \textbf{14} \\
      \midrule
      \textit{Average time (seconds)} & \best{0.53} & \secondbest{0.85} & 1.30 & 2.55 \\
      \bottomrule
    \end{tabularx}

    \caption{Generation time comparison against other methods and amount of LoRAs / Attributes. 
             $^*$Gemini times represent API response duration. }
    \label{tab:time_comparison}
\end{table}

    \begin{paragraph}*{CLIP-Based Evaluation}
    We evaluate CLIP~\cite{radford2021clip} Image Similarity (Figure \ref{fig:clip_image_similarity}) over four facial attributes (age, makeup, anger, curly hair).
    Image Similarity is the cosine between the original and edited image embeddings, and therefore measures structural and identity fidelity. 
    InstructPix2Pix varries a lot also visible in qualitative results.
    Gemini and CharGen maintain consistently high image similarity ($0.85$-$0.90$). 
    Our method matches Gemini image similarity ($0.85$-$0.90$).
    \end{paragraph}

    \begin{figure}
    \centering
    \includegraphics[width=\columnwidth]{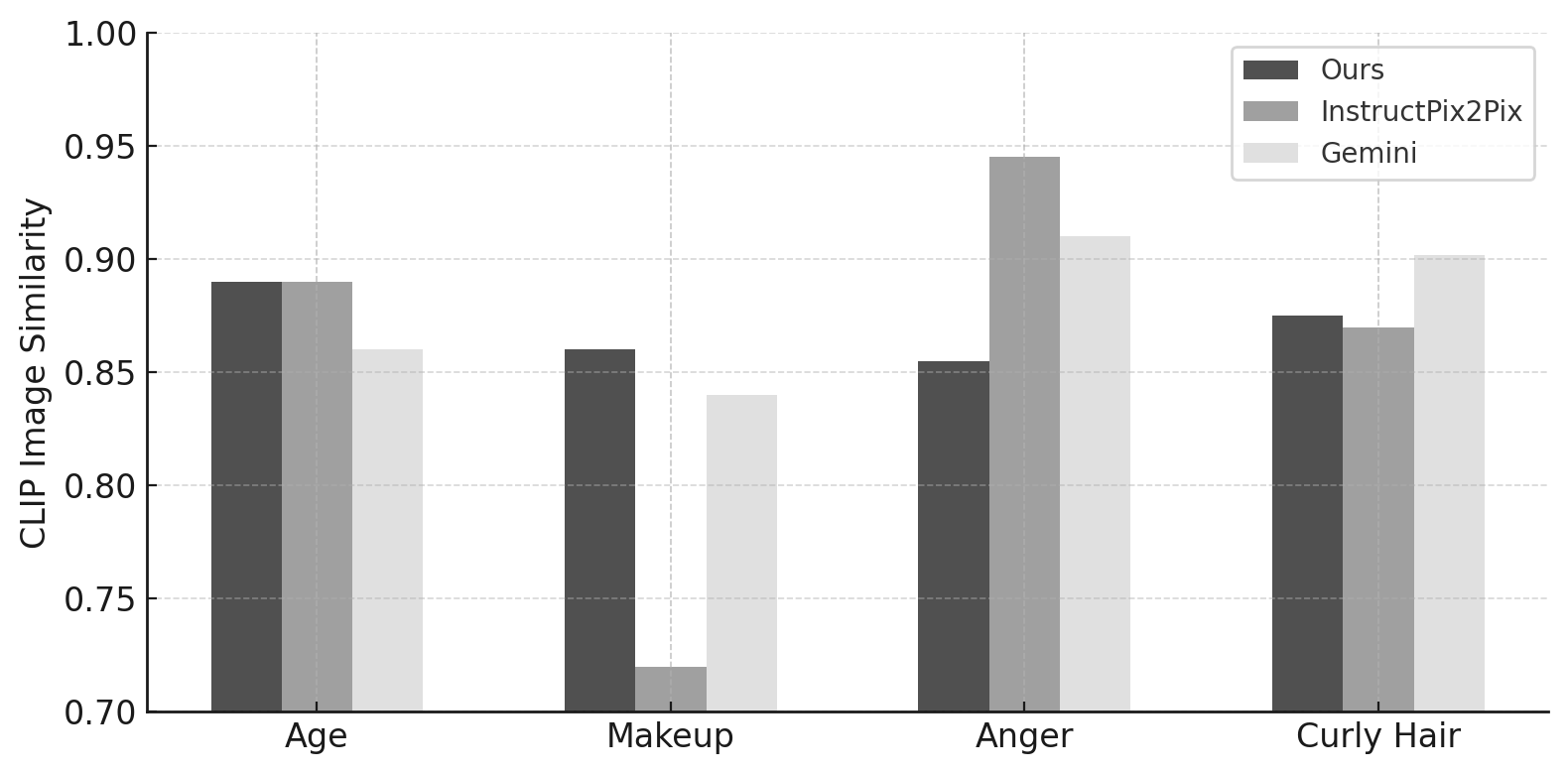}
    \caption{CLIP Image Similarity evaluates the structural fidelity of each editing method to the original input across four attributes (age, makeup, anger, curly hair). Higher values indicate stronger preservation of the input. Results are averaged over ten edited images per attribute.}
    \label{fig:clip_image_similarity}
  \end{figure}

    \begin{paragraph}*{Refinement Step Metrics}
    Standard Stable Diffusion achieves the best perceptual similarity (LPIPS: $0.13$), while our Repair Slider provides slightly improved PSNR and SSIM scores (Table \ref{tab:refinement_metrics}).
    ControlNet produces detailed outputs but significantly alters image structure, confirming our qualitative observations.
    This refinement task is inherently challenging: we remove information from the ground truth image through StreamDiffusion processing, then attempt to generatively reconstruct details without any explicit guidance toward the original ground truth. While our methods don't achieve the same metrics as the unrefined baseline, they successfully add detail information without the excessive structural deviations exhibited by ControlNet.
    \end{paragraph}

    \begin{table}[t]
    \centering
    \scriptsize
    \setlength{\tabcolsep}{4pt} %

    \begin{tabularx}{\linewidth}{l *{4}{>{\centering\arraybackslash}X}}
      \toprule
      \textbf{Metric} & \textbf{ControlNet} & \textbf{SD} & \textbf{SD\,+\,Repair} & \textbf{Unrefined} \\
      \midrule
      LPIPS $\downarrow$ & 0.14 $\pm$ 0.03 & \best{0.13} $\pm$ \best{0.03} & \secondbest{0.14} $\pm$ \secondbest{0.02} & 0.18 $\pm$ 0.05 \\
      PSNR $\uparrow$ & 20.00 $\pm$ 2.12 & \secondbest{22.73} $\pm$ \secondbest{1.19} & \best{22.79} $\pm$ \best{1.21} & 23.69 $\pm$ 1.17 \\
      SSIM $\uparrow$ & 0.59 $\pm$ 0.06 & \secondbest{0.64} $\pm$ \secondbest{0.05} & \best{0.65} $\pm$ \best{0.05} & 0.69 $\pm$ 0.05 \\
      \bottomrule
    \end{tabularx}

    \caption{Quantitative comparison of refinement methods against ground truth. 
    The "Unrefined" baseline represents ground truth processed through StreamDiffusion once (Figure~\ref{fig:refinement_comparison}), which naturally achieves better metrics as it remains closer to the original distribution.}
    \label{tab:refinement_metrics}
\end{table}

    \subsection{User Study}
    We conduct a user study with 35 participants to validate perceptual quality across different editing scenarios.
    Participants evaluate both single-attribute modifications and multi-attribute editing tasks, see Table \ref{tab:user_study} and find more details in the supplementary.

    \begin{paragraph}*{Single Attribute Preferences}
    Gemini receives the highest overall preference, followed by CharGen, with InstructPix2Pix ranking lowest.
    While Gemini optimizes for maximal attribute transformations, CharGen prioritizes subtle, continuous control. 
    Our sliders are trained for fine-grained adjustments rather than dramatic changes, and the training pairs may not capture the full range of attribute variations that Gemini's large-scale training enables. 
    This trade-off reflects our focus on precision and identity preservation over transformation strength.
    \end{paragraph}

    \begin{table}[t]
    \centering
    \scriptsize
    \setlength{\tabcolsep}{4pt} %

    \begin{tabularx}{\linewidth}{l *{3}{>{\centering\arraybackslash}X}}
      \toprule
      \textbf{Editing} & & & \\[-1.1em]
      & \textbf{Gemini} & \textbf{Ours} & \textbf{IP2P} \\
      \midrule
      \textit{Single Attribute}      & \best{41} & \secondbest{31} & 28 \\
      \textit{Multiple Attribute} & 12 & \best{76} & 12 \\
      \midrule
      \midrule
      \textbf{Refinement} & & & \\[-1.1em]
      & \textbf{ControlNet} & \textbf{SD\,+\,Repair} & \textbf{SD} \\
      \midrule
      \textit{Level of detail} & \best{73} & \secondbest{15} & 12 \\
      \textit{Closest to input}      & 2 & \secondbest{43} & \best{55} \\
      \bottomrule
    \end{tabularx}

    \caption{User-study preference (\%, $N\!=\!35$). Showing the editing preference and refinement preference of the participants.}
    \label{tab:user_study}
\end{table}

    \begin{paragraph}*{Multi-Attribute Editing}
    Users demonstrate a strong preference for CharGen in multi-attribute scenarios.
    This validates our approach's effectiveness for complex editing tasks requiring simultaneous attribute adjustments.
    The results confirm that CharGen excels at fine-grained, multi-attribute control rather than large-scale transformations.
    \end{paragraph}

    \begin{paragraph}*{Refinement Step Assessment}
    Users prefer ControlNet's detail level but reject its results when shown alongside input images.
    This confirms the quantitative findings that ControlNet produces detailed but structurally inconsistent outputs.
    Our Repair Slider approach receives consistent positive feedback for maintaining both detail and structural fidelity.
    \end{paragraph}

    The experimental results demonstrate CharGen's effectiveness for interactive character editing, particularly in scenarios requiring fine-grained, multi-attribute control.
    While existing methods excel at large-scale transformations, CharGen provides the precision and speed necessary for iterative editing workflows.
    A comprehensive video demonstrating continuous control capabilities, and the user study are provided in the supplementary.
\end{section}

\begin{section}{Conclusion}
    \label{sec:conclusion}

    We presented CharGen, a fluent character editing system that combines attribute-specific Concept Sliders with StreamDiffusion's accelerated pipeline to enable interactive portrait modification.
    Our approach addresses the fundamental trade-offs between fine-grained control, generation speed, and visual fidelity in diffusion-based editing.

    CharGen achieves two-to-fourfold faster edit turnaround compared to standard diffusion methods while maintaining precise attribute control through pretrained Concept Sliders.
    The lightweight Repair Step effectively restores high-frequency details lost during accelerated sampling, with the Repair Slider approach providing the most consistent detail enhancement.
    Our evaluation demonstrates that CharGen excels at subtle, multi-attribute modifications while preserving character identity, though it shows limitations for complex transformations like extreme age changes.

    The user study validates CharGen's effectiveness for interactive workflows, with participants preferring our approach for sequential editing tasks over text-based alternatives.
    However, the study also reveals the inherent challenge of balancing edit strength with identity preservation, particularly for the ControlNet-based repair method.

    While CharGen provides valuable tools for creative character editing, it also raises important ethical concerns. The system's ability to manipulate facial attributes could potentially be misused for creating deepfakes or misleading content. Additionally, the system's performance may exhibit biases across different demographic groups due to the underlying training data distribution. We emphasize that CharGen should be used responsibly, with clear disclosure when edited images are presented.

    Future work could extend CharGen's attribute-specific approach to broader image editing domains beyond facial features.
    Key research directions include developing strategies to mitigate unwanted interactions between Concept Sliders, improving training methodologies for attributes with discrete rather than continuous variations, and exploring more sophisticated LoRA integration techniques for enhanced detail generation.

    CharGen contributes to the broader field of controllable generative editing by demonstrating how interactive performance can be achieved without sacrificing the precision and fidelity required for professional character editing workflows.
\end{section}

\vspace{-0.25cm}
\begin{section}{Acknowledgements}
    \label{sec:acknowledgements}

    Funded by the Deutsche Forschungsgemeinschaft (DFG, German Research Foundation) under Germany's Excellence Strategy – EXC number 2064/1 – Project number 390727645.
    This work was supported by the German Research Foundation (DFG): SFB 1233, Robust Vision: Inference Principles and Neural Mechanisms, TP 02, project number: 276693517.
    This work was supported by the Tübingen AI Center.
    The authors thank the International Max Planck Research School for Intelligent Systems (IMPRS-IS) for supporting Jan-Niklas Dihlmann.
\end{section}

\vspace{-0.25cm}
\bibliographystyle{eg-alpha-doi}
\bibliography{bibliography}

\end{document}